\documentclass[conference,a4paper]{IEEEtran}
\usepackage{cite}
\usepackage{amsmath,amssymb,amsfonts}
\usepackage{graphicx}
\usepackage{textcomp}
\usepackage{xcolor}
\usepackage{booktabs}
\usepackage{url}
\usepackage{hyperref}
\usepackage{orcidlink}
\usepackage{balance}  
\hypersetup{
  hidelinks,
  pdftitle={Missing Light, Misleading Confidence: Optical-Module Readout Loss and Signal Reliability in Neutrino Telescopes},
  pdfauthor={Thammarat Yawisit and Pittaya Pannil},
  pdfsubject={Preprint submitted to the 2027 IEEE Sensors Applications Symposium (SAS)},
  pdfkeywords={neutrino telescopes, optical sensors, readout loss, data quality, selective acceptance}
}
\makeatletter
\def\ps@arxivfirst{%
  \def\@oddhead{}%
  \def\@evenhead{}%
  \def\@oddfoot{\parbox[b]{\textwidth}{\centering\footnotesize
    Preprint. Submitted to the 2027 IEEE Sensors Applications Symposium (SAS).\\[2pt]
    This work has been submitted to the IEEE for possible publication.\\
    Copyright may be transferred without notice, after which this version may no longer be accessible.}}%
  \let\@evenfoot\@oddfoot
}
\makeatother
\graphicspath{{figures/}}
\newtheorem{proposition}{Proposition}

\begin{document}

\title{Missing Light, Misleading Confidence: Optical-Module Readout Loss and Signal Reliability in Neutrino Telescopes}

\author{\IEEEauthorblockN{Thammarat Yawisit\,\orcidlink{0009-0006-9298-8862}}
\IEEEauthorblockA{\textit{Department of Instrumentation and Control Engineering}\\
\textit{School of Engineering}\\
\textit{King Mongkut's Institute of Technology Ladkrabang}\\
Bangkok 10520, Thailand\\
65010454@kmitl.ac.th (corresponding author)}
\and
\IEEEauthorblockN{Pittaya Pannil\,\orcidlink{0000-0001-6082-3058}}
\IEEEauthorblockA{\textit{Department of Instrumentation and Control Engineering}\\
\textit{School of Engineering}\\
\textit{King Mongkut's Institute of Technology Ladkrabang}\\
Bangkok 10520, Thailand\\
pittaya.pa@kmitl.ac.th}
}

\maketitle
\thispagestyle{arxivfirst}

\begin{abstract}
Optical neutrino telescopes reduce the photomultiplier signals of an optical module or string segment to compact multichannel summaries, and readout loss can make these summaries unreliable. We asked whether availability-based quality gates, which accept a summary when enough of the readout was live, can certify a summary of inter-channel statistical consistency (called coherence here; not optical coherence) under readout loss, and where they fail. We used public IceCube pulses (9,484 string records) and the released hits of KM3NeT event KM3-230213A with controlled readout loss, together with a replay of real KM3NeT readout flags. In a two-by-two design we compared a published recursive support state with direct mask support, and calibrated every threshold on development records. The support state is a function of the mask and added no information. On IceCube the support-gated rule retained 0.16 percentage points more usable summaries, an effect of the window at which support was read that reversed sign at a looser tolerance; on KM3NeT it accepted nothing. The gates accepted only records with nearly intact masks, including records whose charge had been depleted by signal-dependent loss that the mask does not record. Pulse loss during live acquisition and timing jitter passed every availability gate, which then accepted summaries of which up to 66\% were unreliable. Availability flags are therefore necessary but not sufficient, and signal-based checks are needed. This work uses public data and is not an IceCube or KM3NeT Collaboration result.
\end{abstract}


\section{Introduction}
In an optical neutrino telescope, Cherenkov photons release photoelectrons in a photomultiplier tube (PMT), and the front-end electronics turn each pulse into a time and a charge, as in the IceCube digital optical module (DOM)~\cite{icecubedaq,icecubeinst}, or a time and a time over threshold (ToT), as in the 31-PMT KM3NeT module~\cite{km3readout,km3mpmt}. The resulting hits are combined across the PMTs of one module or the modules of one string segment. Statistics of inter-channel consistency enter local-coincidence and hit-cleaning conditions and module monitoring~\cite{icecubeinst}, and multi-PMT coincidences within a module drive the KM3NeT supernova trigger~\cite{km3sn}. Readout is not always complete: the KM3NeT high-rate veto (HRV) drops a PMT's hits for a 100~ms timeslice and a FIFO flag marks possible truncation~\cite{km3readout,km3sn}, so the number of active PMTs varies in time. Single-PMT strings (Baikal-GVD~\cite{baikal2020}, the proposed HUNT~\cite{hunt2025}) and multi-PMT modules (P-ONE~\cite{pone,ponecal2026}) face the same question.

We ask: \emph{can availability-based quality gates certify multichannel summaries of optical-module data under readout loss, and where do they fail?} A gate accepts a summary when an availability score passes a threshold calibrated to a target false-acceptance rate. This is selective acceptance~\cite{elyaniv,geifman}, with risk-controlling calibration methods~\cite{crc,ltt,jincandes}. Two gates are compared: the direct pair support of the availability mask, and the recursive observation-support state $Q$ of the time-aligned multichannel coherence estimator TAMCE~\cite{tamce}, whose authors noted that $Q$ largely encodes the mask. Mask-based gates can only see loss that the mask records; unrecorded or signal-dependent loss falls outside the missing-at-random setting~\cite{rubin,littlerubin}.

Contributions: a $2\times2$ estimator-by-gate analysis on measured IceCube and KM3NeT data with full coverage and risk reporting; realised loss severity per mechanism; a replay of real KM3NeT readout flags; and a demonstration that pulse loss and timing jitter pass every availability gate. Here coherence means statistical consistency between channels only; no photons or hits are recovered. This work uses publicly released data and is not an IceCube or KM3NeT Collaboration result.

\begin{figure*}[!t]
\centering
\includegraphics[width=0.86\textwidth]{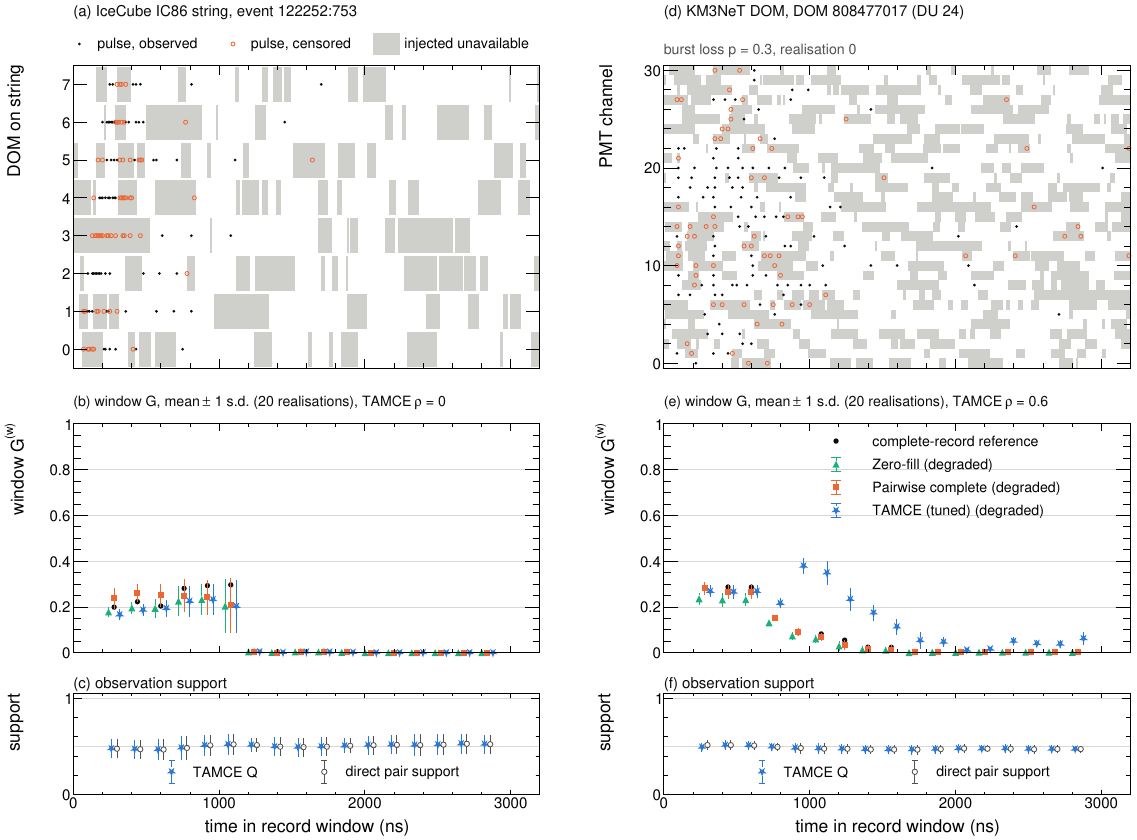}
\caption{One measured test record per telescope with injected burst loss at requested fraction 0.3. Left: IceCube IC86 string (eight adjacent DOMs; 90th-percentile record by pulse count). Right: KM3NeT DOM 808477017 of event KM3-230213A (median test DOM). (a, d) One realisation: pulses (filled circles), unavailable intervals (grey), removed pulses (open circles). (b, e) Per-window coherence $G^{(w)}$: complete-record reference (filled circles) and degraded-record estimates by zero-fill (triangles), pairwise-complete correlation (squares) and tuned TAMCE (stars), mean $\pm1$ s.d. over 20 loss realisations. (c, f) TAMCE support $Q^{(w)}$ (stars) and direct pair support $S^{(w)}$ (open circles).}
\label{fig:example}
\end{figure*}

\section{Data}
\textbf{IceCube IC86 (measured).} We used the two files of the IceCube public data page~\cite{icecubeodf}: the LargeNCh sample of run 122252 (1,329 events) and the neutrino sample of run 121585 (71 events), 1.80 million calibrated pulses. Our reader of the IceCube Open Data Format reproduced the official \texttt{odf\_tools} reader on all 1,400 events with zero difference. For each event and string, the record is the window of eight adjacent DOMs (17~m spacing, no larger gaps) with the largest charge, as 320 bins of 10~ns of summed charge starting 100~ns before the record's 2nd-percentile pulse time. A record needed at least 20 pulses inside its window, on at least four DOMs; an earlier version of our builder applied this rule before windowing, and the 790 records it wrongly admitted are excluded. This gave 9,484 records from 1,331 events, split by event (4,797 dev; 4,687 test from 667 events). The data carry no DOM-status information, so all DOMs were assumed live.

\textbf{KM3NeT KM3-230213A (measured, illustrative).} The released hits (CC-BY 4.0)~\cite{km3event,km3data} gave one record per DOM with at least 100 merged hits (74 records): 31 PMTs by 320 bins of 10~ns, with the ToT sum divided by the median ToT of all released hits (27~ns) as feature, starting 100~ns before the DOM's 2nd-percentile hit time. The split is by detection unit (DU): 38 dev DOMs, and 36 test DOMs on only 6 DUs. This is an illustrative case study of one selected event, and its intervals describe that event only.

\textbf{Empirical DAQ-mask replay (post hoc).} Three DAQ files of the KM3NeT software test package \texttt{km3net-testdata} 0.4.31~\cite{km3testdata} (detectors 29 and 44, runs 3246, 6633 and 6884) give 18,047 summary frames with per-PMT HRV and FIFO bits; 17\% of frames flag at least one PMT. We drew one flag vector per record and fault seed, from all frames or from flagged frames only, and marked the flagged PMTs unavailable for the whole 3.2~$\mu$s record. Bit $c$ was mapped to PMT channel $c$ of the DOM, following the \texttt{km3io} convention; HRV (hits dropped) and FIFO (hits possibly truncated) were both treated as full loss. The package does not identify the site of these detectors, which differ from the ARCA configuration that recorded KM3-230213A. The replay therefore applies a realistic flag pattern to measured hits; it is not a matched operational validation.

\textbf{Synthetic.} Five configurations generated by our own code from published designs are reported as supporting evidence only (Table~\ref{tab:primary}); they are not collaboration simulations.

\section{Methods}\label{sec:methods}
\begin{figure*}[!t]
\centering
\includegraphics[width=0.86\textwidth]{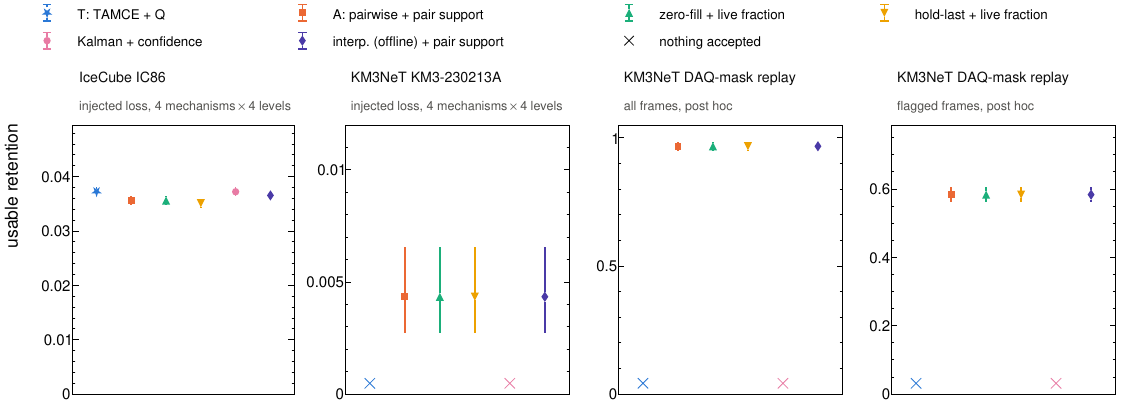}
\caption{Usable retention $U$ at $\alpha=0.10$ on measured test data. Stars: rule T; squares: rule A; up- and down-triangles: zero-fill and hold-last with live fraction; circles: tuned Kalman with its confidence; diamonds: offline interpolation (non-causal). Bars: 95\% cluster bootstrap. $\times$: nothing accepted, so selective risk is undefined. Each panel has its own scale; the two right panels are the DAQ-mask replay (post hoc).}
\label{fig:ret}
\end{figure*}

\begin{figure*}[!t]
\centering
\includegraphics[width=0.78\textwidth]{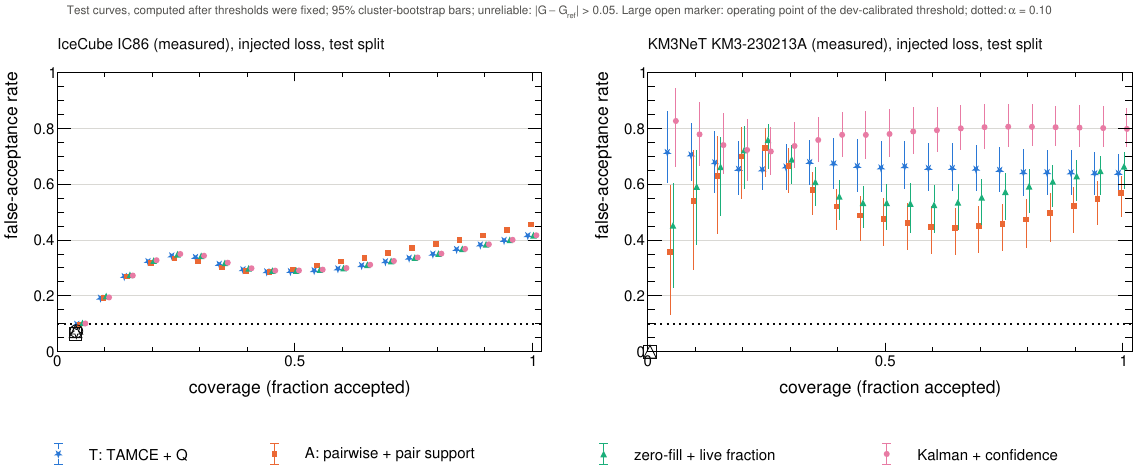}
\caption{Test risk--coverage on measured data, evaluated after the thresholds were fixed: selective risk ($|e|>0.05$) against coverage, with summaries ranked by each rule's score. Markers as in Fig.~\ref{fig:ret}; bars: 95\% cluster bootstrap; large open markers: operating points set by the dev-calibrated thresholds; dotted: $\alpha=0.10$. On KM3NeT, T and Kalman have no operating point because no threshold met $\alpha$ on the dev records.}
\label{fig:rc}
\end{figure*}

\begin{figure*}[p]
\centering
\includegraphics[width=0.86\textwidth]{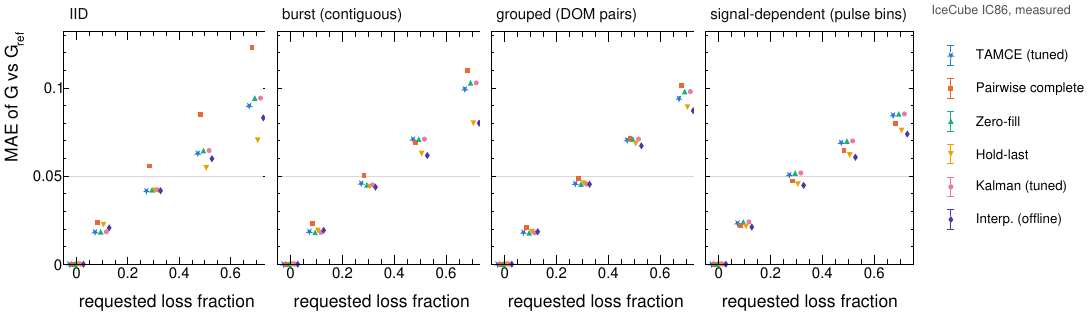}\\[-1pt]
{\footnotesize (a) IceCube IC86 string records, measured (test split, 4,687 records from 667 events)}\\[4pt]
\includegraphics[width=0.86\textwidth]{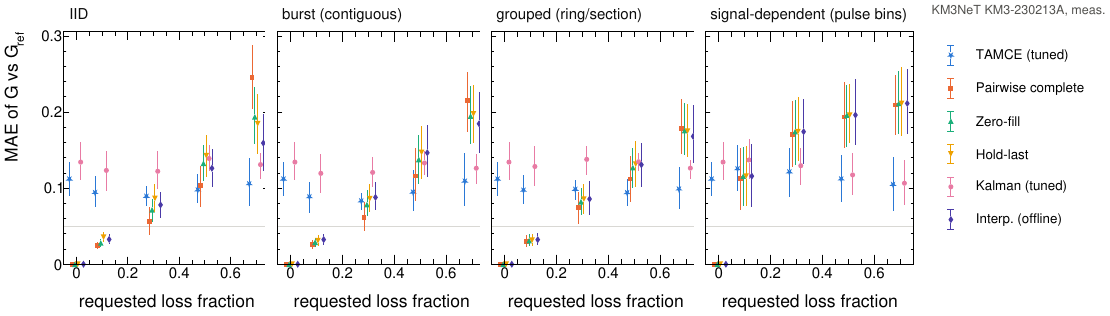}\\[-1pt]
{\footnotesize (b) KM3NeT DOMs of event KM3-230213A, measured (test split, 36 DOMs on 6 detection units)}
\caption{Mean absolute error of $G$ against $G_\text{ref}$ versus requested loss fraction, for the four injected mechanisms. The value at zero loss is each estimator's intrinsic bias. Markers: stars TAMCE, squares pairwise complete, up-triangles zero-fill, down-triangles hold-last, circles Kalman, diamonds offline interpolation; bars: 95\% cluster bootstrap (IceCube events, KM3NeT DUs). Points are offset horizontally for legibility.}
\label{fig:sev}
\vspace{10pt}
\includegraphics[width=0.86\textwidth]{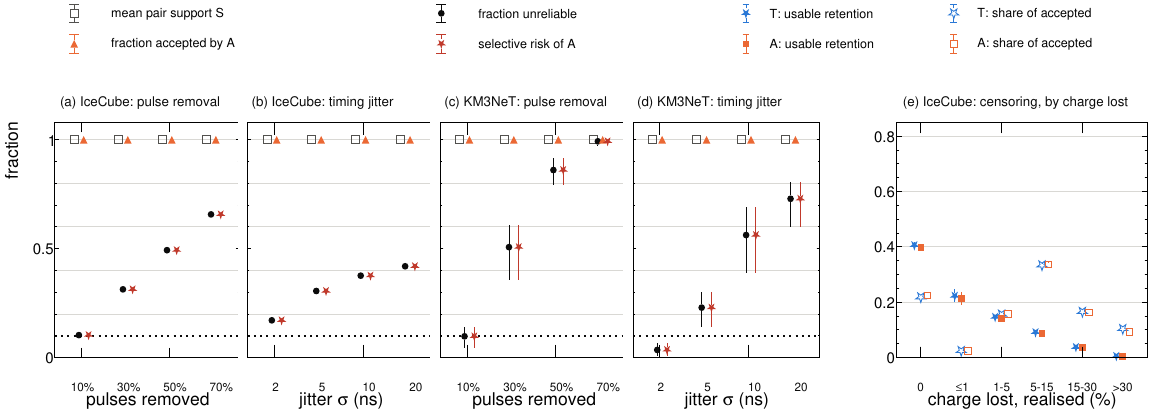}
\caption{Failures invisible to availability (test split, rule A at its dev-calibrated threshold). (a--d) Pulse removal with acquisition live and per-channel timing jitter: mean pair support (open squares), fraction unreliable (filled circles), fraction accepted (triangles) and selective risk among accepted (stars); dotted: $\alpha=0.10$. (e) Rules T (stars) and A (squares) on the injected censoring mechanisms, by realised fraction of charge lost: usable retention (filled) and share of all accepted summaries (open). Bars: 95\% cluster bootstrap.}
\label{fig:inv}
\end{figure*}

\textbf{Summary statistic.} A record has $N$ channels $z_i[n]$, $n=1,\dots,320$, and an availability mask $m_i[n]\in\{0,1\}$; $m=0$ means unavailable (value unknown), whereas $m=1$ with $z=0$ means live without a pulse. Windows $w$ have $W=56$ bins and start at $s\in\{0,16,\dots,256\}$, measured from the record start. Within $w$, $R_{ij}^{(w)}$ is the Pearson correlation of channels $i$ and $j$, and
\begin{equation}
G^{(w)}=\Big(\tbinom{N}{2}^{-1}\textstyle\sum_{i<j}\big(R_{ij}^{(w)}\big)^2\Big)^{1/2},\qquad G=\max_w G^{(w)}.
\end{equation}
The RMS keeps anticorrelated pairs, and the maximum locates the signal window without a trigger time. A pair with a zero-variance channel contributes $R_{ij}=0$; there is no other normalisation. $G_\text{ref}$ is the direct functional on the complete record. The measured records were not re-aligned, because no per-channel offsets are released.

\textbf{Estimators.} Pairwise-complete correlation uses, for each pair, only jointly available samples, and sets $R_{ij}=0$ below 8 joint samples. TAMCE~\cite{tamce} runs, from $b_i=q_i=0$,
\begin{equation}
\begin{gathered}
b_i\leftarrow\rho b_i+(1-\rho)m_iz_i,\qquad q_i\leftarrow\rho q_i+(1-\rho)m_i,\\
\hat x_i=b_i/\max(q_i,0.02),
\end{gathered}
\end{equation}
and correlates $Y_i=\sqrt{q_i}\,(\hat x_i-\mu_i)$, with $\mu_i$ the $q$-weighted window mean, so that unavailable samples get zero weight. Its support is $Q^{(w)}=\binom{N}{2}^{-1}\sum_{i<j}W^{-1}\sum_{n\in w}q_iq_j$. The direct pair support $S^{(w)}$ is the same expression with $m$ in place of $q$. Each score is read at the window $w^*$ where the rule's own estimate of $G$ is largest, so $Q^*=Q^{(w^*(z,m))}$ depends on the signal only through $w^*$. The memory $\rho$ was tuned on dev records by RMSE to $G_\text{ref}$, giving $\rho=0$ on IceCube and $\rho=0.6$ on KM3NeT. Zero-fill, hold-last, a diagonal Kalman filter~\cite{kalman,sinopoli} (confidence $1/(1+P)$ read like $Q$) and offline interpolation are further baselines.

\begin{proposition}
$q_i[n]=(1-\rho)\sum_{k\ge0}\rho^k m_i[n-k]$, so $Q^{(w)}$ is a function of the mask history only. For $\rho=0$, $q_i=m_i$ and $Q^{(w)}=S^{(w)}$.
\end{proposition}

\textbf{Readout loss.} Requested fractions $p\in\{0.1,0.3,0.5,0.7\}$ under four mechanisms: IID, burst (mean gap 10 bins), grouped (KM3NeT PMT rings, IceCube DOM pairs; mean gap 20 bins) and signal-dependent (pulse-bearing bins only, probability $\propto$ pulse size), each with four fault seeds per record and condition. Two stresses leave the mask at 1: pulse removal (10--70\%) and per-channel timing offsets (2--20~ns).

\textbf{Endpoints.} A summary is usable if $|e|=|G-G_\text{ref}|\le\tau=0.05$. A gate accepts when its score is at least $t$. Over $n$ test summaries with $n_a$ accepted and $n_{au}$ of them unusable, coverage is $C=n_a/n$, selective risk $R=n_{au}/n_a$ (undefined if $n_a=0$), and usable retention $U=(n_a-n_{au})/n=C(1-R)$, the primary endpoint. We also report the self-reference error $G_j(\text{degraded})-G_j(\text{complete})$, which measures stability rather than fidelity. The threshold is the loosest $t$ with dev $R\le\alpha=0.10$, applied unchanged to test; scores are rounded to 12 decimals so that equal supports form one tie block. This empirical calibration carries no finite-sample guarantee; a Clopper--Pearson upper bound~\cite{geifman} was used as a sensitivity analysis. The hypothesis (T retains more than A), $\tau$, $\alpha$, grids and inference were fixed in advance but not publicly registered; analyses added later are marked post hoc. Inference uses a 95\% cluster bootstrap with $B=2000$~\cite{efron} over IceCube events, KM3NeT DUs and synthetic records, with Holm correction~\cite{holm} across the synthetic configurations.

\section{Results}

\begin{table*}[!t]
\caption{Primary test ($\tau=0.05$, $\alpha=0.10$, test split): accepted count $n_a$, coverage $C$, selective risk $R$ and usable retention $U$ for rule T (TAMCE gated by $Q$) and rule A (pairwise-complete gated by $S$). N/A: nothing accepted. CI: 95\% cluster bootstrap of the paired difference.}
\label{tab:primary}
\centering\footnotesize\setlength{\tabcolsep}{4pt}
\begin{tabular}{@{}lrrcccrcccc@{}}
\toprule
& & \multicolumn{4}{c}{T: TAMCE + $Q$} & \multicolumn{4}{c}{A: pairwise + $S$} & \\
\cmidrule(lr){3-6}\cmidrule(lr){7-10}
Dataset & $n$ & $n_a$ & $C$ & $R$ & $U$ & $n_a$ & $C$ & $R$ & $U$ & T$-$A [95\% CI]\\
\midrule
IceCube IC86 (measured) & 299,968 & 12,017 & 0.040 & 0.071 & 0.037 & 11,433 & 0.038 & 0.065 & 0.036 & $+0.0016$ [$+0.0014$, $+0.0018$]\\
KM3-230213A (measured) & 2,304 & 0 & 0 & N/A & 0 & 10 & 0.004 & 0.000 & 0.004 & $-0.004$ [$-0.007$, $-0.003$]\\
\quad DAQ-mask replay, all frames$^\dagger$ & 576 & 0 & 0 & N/A & 0 & 576 & 1.000 & 0.033 & 0.967 & $-0.967$ [$-0.984$, $-0.952$]\\
\quad DAQ-mask replay, flagged frames$^\dagger$ & 576 & 0 & 0 & N/A & 0 & 349 & 0.606 & 0.037 & 0.583 & $-0.583$ [$-0.606$, $-0.561$]\\
\midrule
IceCube-like (synthetic) & 4,800 & 0 & 0 & N/A & 0 & 597 & 0.124 & 0.096 & 0.112 & $-0.112$ [$-0.124$, $-0.101$]\\
KM3NeT-like DOM (synthetic) & 4,800 & 0 & 0 & N/A & 0 & 194 & 0.040 & 0.062 & 0.038 & $-0.038$ [$-0.045$, $-0.031$]\\
Baikal-GVD-like (synthetic) & 4,800 & 834 & 0.174 & 0.094 & 0.158 & 912 & 0.190 & 0.094 & 0.172 & $-0.015$ [$-0.020$, $-0.010$]\\
HUNT-like (synthetic) & 4,800 & 805 & 0.168 & 0.128 & 0.146 & 1,030 & 0.215 & 0.132 & 0.186 & $-0.040$ [$-0.047$, $-0.033$]\\
P-ONE-like (synthetic) & 4,800 & 0 & 0 & N/A & 0 & 432 & 0.090 & 0.134 & 0.078 & $-0.078$ [$-0.087$, $-0.069$]\\
\bottomrule
\multicolumn{11}{@{}l@{}}{\scriptsize $^\dagger$Real KM3NeT readout flags from other runs instead of injected loss; added post hoc. Synthetic rows: Holm-adjusted $p<0.001$.}
\end{tabular}
\end{table*}

\textbf{Primary test} (Table~\ref{tab:primary}, Fig.~\ref{fig:ret}). On IceCube rule T retained $0.0016$ [$0.0014$, $0.0018$] more usable summaries than A, at $U\approx0.04$ for both and $R$ below $\alpha$. On KM3NeT no $Q$ threshold met $\alpha$ on dev records, so T accepted nothing, and A accepted 10 of 2,304 test summaries. Every leave-one-DU-out estimate of T$-$A stayed between $-0.0048$ and $-0.0036$, and the DU-level sign-flip test gave $p=0.031$, the smallest value possible with six DUs. In the DAQ-mask replay, A kept $U=0.97$ (all frames) and $0.58$ (flagged frames) at $R\le0.04$; T accepted nothing. Across the rules and datasets tested we found no incremental benefit of $Q$ over direct availability support.

\begin{table}[t]
\caption{$2\times2$ estimator-by-gate design: usable retention $U$ on the measured test splits, and main effects in units of $10^{-4}$ (mean paired differences, 95\% cluster bootstrap).}
\label{tab:fact}
\centering\footnotesize\setlength{\tabcolsep}{3pt}
\begin{tabular}{@{}lcccc@{}}
\toprule
& \multicolumn{2}{c}{IceCube IC86} & \multicolumn{2}{c}{KM3-230213A}\\
\cmidrule(lr){2-3}\cmidrule(lr){4-5}
Estimator & gate $S$ & gate $Q$ & gate $S$ & gate $Q$\\
\midrule
Pairwise complete & 0.0357 (A) & 0.0373 & 0.0043 (A) & 0.0013\\
TAMCE (tuned) & 0.0372 & 0.0372 (T) & 0 & 0 (T)\\
\midrule
Estimator effect & \multicolumn{2}{c}{$+8$ [$+7$, $+8$]} & \multicolumn{2}{c}{$-28$ [$-46$, $-17$]}\\
Gate effect & \multicolumn{2}{c}{$+8$ [$+7$, $+9$]} & \multicolumn{2}{c}{$-15$ [$-26$, $-8$]}\\
Interaction & \multicolumn{2}{c}{$-17$ [$-18$, $-15$]} & \multicolumn{2}{c}{$+30$ [$+15$, $+52$]}\\
\bottomrule
\end{tabular}
\end{table}

\textbf{Gate against estimator} (Table~\ref{tab:fact}). On IceCube, with $\rho=0$, $Q$ and $S$ gave identical acceptance sets for TAMCE, as the proposition requires. Gating pairwise $G$ by $Q$ beat A by $0.0016$ only because $Q$ is read at TAMCE's window rather than pairwise's; the gate effect is a window-reading effect. On KM3NeT both main effects were negative. Estimator fidelity and gate quality also separate: with no loss, tuned TAMCE ($\rho=0.6$) differs from $G_\text{ref}$ by a mean absolute error of 0.112, above $\tau$, while it reproduces its own complete-record value exactly. Under the self-reference endpoint T retained 0.010 against 0.004 for A on KM3NeT (difference $+0.006$ [$-0.001$, $+0.010$]), so T loses on fidelity to the direct reference rather than on stability. Its memory cost fidelity wherever it was kept: IceCube dev RMSE rose from 0.089 at $\rho=0$ to 0.398 at $\rho=0.75$, and with $\rho=0$ on KM3NeT, T reproduced A exactly ($U=0.0043$).

\textbf{What the gates accept.} IceCube thresholds sat near full support (0.992). Of T's accepted summaries, 64\% came from signal-dependent loss, 35\% from grouped loss, 1\% from burst loss and none from IID loss. Signal-dependent loss hides from the mask: at $p=0.3$ it removed 0.7\% of the channel--time cells but 43\% of the charge. Usable retention was 0.23 when at most 5\% of the charge was lost and 0.023 above that (Fig.~\ref{fig:inv}e), where the selective risk of T was 0.11, above $\alpha$; 60\% of T's accepted summaries lay in that range. The gates thus keep records with nearly intact masks, not necessarily intact signals.

\textbf{Invisible failures} (Fig.~\ref{fig:inv}a--d). Pulse removal and jitter left the mask, and so $S$ and $Q$, at 1.0. Rule A therefore accepted every stressed summary. On IceCube its selective risk rose from 0.10 to 0.66 as 10--70\% of pulses were removed, and to 0.42 at 20~ns of jitter; on KM3NeT it reached 0.99 and 0.73. Per-hit jitter (risk 0.45 at 20~ns), common-mode offsets (0.23) and slow per-channel drift (0.20) were likewise accepted in full.

\textbf{Sensitivity.} The IceCube difference was not robust: it was $+0.0010$ at $\tau=0.025$, $+0.0009$ at $\alpha=0.20$ and $+0.0010$ at a relative tolerance $|e|/\max(G_\text{ref},0.05)\le0.1$, but $-0.0017$ at $\tau=0.10$ and $-0.012$ at a relative tolerance of 0.2. Clopper--Pearson thresholds left IceCube unchanged and made both KM3NeT rules accept nothing; calibrating without the 60 tuning records changed nothing. Aggregating windows by their mean or median made the summary degenerate (the median $G_\text{ref}$ is 0 for 63\% of records), and reading one fixed window removed the IceCube difference ($-0.0001$ [$-0.0002$, $0.0000$]).

\section{Discussion and limitations}
$Q$ is a deterministic function of the mask history, so as a gate it can at most reproduce availability information; differences between rules come from the estimator and from the window at which support is read. The gates certified only nearly intact records and accepted unreliable summaries without limit under unrecorded loss. Availability flags are therefore necessary but not sufficient, and a usable quality flag needs signal-based checks such as pulse-count sufficiency, timing-residual or lag consistency and binning stability, which we evaluate in a companion journal study. In a post hoc first test, pulse-count and stability scores cut the accepted share of pulse-removal and jitter summaries on IceCube from 100\% to 5\% and 14\%, but their selective risk stayed above $\alpha$. The tolerance $\tau=0.05$ is a statistical choice, about one third of the interquartile range of $G_\text{ref}$ on KM3NeT DOMs, not a physics requirement. Limitations: $G_\text{ref}$ is computed from released records that may already contain unrecorded loss (HRV or FIFO in the KM3NeT event, unknown DOM status in IceCube); errors are relative to the released record and understate the departure from complete data, while the comparison between rules shares that reference. The KM3NeT evidence is one event on six test DUs; the flag replay uses other detectors and was added post hoc; injected masks are stress models, not detector deadtime models. Nothing here validates any method on operating telescopes.

\section{Conclusion}
On measured IceCube and KM3NeT data, availability gates, including the TAMCE support state, accepted only records with nearly intact masks, and $Q$ added no information beyond the mask. Pulse loss and timing errors passed every availability gate. Quality flags for multichannel summaries in neutrino telescopes need signal-based checks in addition to availability.

\section*{Data and code availability}
IceCube open data: \url{https://icecube.umd.edu/PublicData/}. KM3NeT event hits: \url{https://github.com/KM3NeT/KM3-230213A-data} (CC-BY 4.0). KM3NeT DAQ test files: \texttt{km3net-testdata} 0.4.31 (MIT). TAMCE: arXiv:2512.20678v3. Code (version 1.1.0, including the ODF reader and its check against \texttt{odf\_tools}), configurations, raw results and logs are provided as the reproducibility archive accompanying this submission and will be deposited in a public repository with a DOI before publication.

\balance

\end{document}